\documentclass[iop,apj]{emulateapj}
\usepackage{amsmath}
\def\lsim{\mathrel{\rlap{\lower4pt\hbox{\hskip1pt$\sim$}}    \raise1pt\hbox{$<$}}} 
\def\aj{AJ}
\def\apj{ApJ}

\def\apjl{ApJL}
\def\mnras{MNRAS}
\def\aap{A\&A}
\def\araa{ARA\&A}

\shorttitle{The nearest ultra diffuse galaxy}
\shortauthors{Trujillo  et al.}

\begin{document}

\title{The nearest ultra diffuse galaxy: UGC2162}

\author{Ignacio Trujillo\altaffilmark{1,2}, Javier Roman\altaffilmark{1,2}, Mercedes Filho\altaffilmark{1,2} and Jorge S\'anchez Almeida\altaffilmark{1,2}}
  
\email{trujillo@iac.es}

\altaffiltext{1}{Instituto de Astrof\'isica de Canarias, Calle V\'ia Lactea, La Laguna, Tenerife, Spain}
\altaffiltext{2}{University of La Laguna. Avda. Astrof\'isico Fco. S\'anchez, La Laguna, Tenerife, Spain}

\begin{abstract}

 We describe the structural, stellar population and gas properties of the nearest Ultra Diffuse Galaxy (UDG) discovered so far:
 UGC2162 (z=0.00392; R$_{e,g}$=1.7$(\pm$0.2) kpc; $\mu_g(0)$=24.4$\pm$0.1 mag/arcsec$^2$; g-i=0.33$\pm$0.02). This galaxy, located at
 a distance of 12.3($\pm$1.7) Mpc, is a member of the M77 group. UGC2162 has a stellar mass of $\sim$2($^{+2}_{-1}$)$\times$10$^7$
 M$_\odot$ and is embedded within a cloud of HI gas $\sim$10 times more massive: $\sim$1.9($\pm$0.6)$\times$10$^8$ M$_\odot$. Using
 the width of its HI line as a dynamical proxy, the enclosed mass within the inner R$\sim$5 kpc is $\sim$4.6($\pm$0.8)$\times$10$^9$
 M$_\odot$ (i.e. M/L$\sim$200). The estimated virial mass from the cumulative mass curve is $\sim$8($\pm$2)$\times$10$^{10}$
 M$_\odot$.  Ultra-deep imaging  from the IAC Stripe82 Legacy Project show that the galaxy is irregular and has many star forming
 knots, with a gas-phase metallicity around one-third of the solar value. Its estimated Star Formation Rate (SFR) is $\sim$0.01 M$_\odot$/yr.
 This SFR would double the stellar mass of the object in $\sim$2 Gyr. If the object were to stop forming stars at this moment, after
 a passive evolution, its surface brightness would become extremely faint: $\mu_g(0)$$\sim$27 mag/arcsec$^2$ and its size would
 remain large R$_{e,g}$$\sim$ 1.8 kpc. Such faintness would make it almost undetectable to most present-day surveys. This suggests
 that there could be an important population of M$_{\star}$$\sim$10$^7$ M$_\odot$ "dark galaxies" in rich environments (depleted of
 HI gas) waiting to be discovered by current and future ultra-deep surveys.

 \end{abstract}


\section{Introduction}

In the last few years there has been a renewed interest in the study of extended low-surface brightness galaxies
\citep[][]{1988ApJ...330..634I,1991ApJ...376..404B,1997AJ....114..635D,2006ApJ...651..822C}. The discovery of
dozens of these objects in the Coma Cluster \citep[coined UDGs by][]{2015ApJ...798L..45V} has been followed by a
large number of detections in other clusters
\citep[][]{2015ApJ...807L...2K,2015ApJ...809L..21M,2015ApJ...813L..15M,2016A&A...590A..20V,2016arXiv160303494R},
groups \citep[][]{2016arXiv161008980R,2016A&A...596A..23S,2016arXiv161001609M} and in the field
\citep[][]{2016AJ....151...96M}. The low stellar mass (10$^7$-10$^8$ M$_\odot$) of these objects together with
their large size (R$_e>$1.5 kpc) have opened a number of questions about the ultimate nature of these galaxies: 
are UDGs ``failed'' galaxies  \citep[i.e. do they inhabit dark matter halos larger than those expected according to their stellar mass content;][]{2015ApJ...798L..45V,2016ApJ...830...23B}? What is the role of environment?  Are the properties of UDGs produced by their interaction with dense environments
\citep[][]{2015MNRAS.452..937Y}? Are they simply the high-spin tail of normal dwarf galaxies
\citep[][]{2016MNRAS.459L..51A}? Are UDGs produced by feedback-driven gas outflows and subsequent dark matter and
stellar expansion \citep[][]{2017MNRAS.466L...1D}?

Observations indicate that UDGs are a heterogeneous population of dwarf galaxies. Some of them are relatively red (g-i$\sim$0.8),
have spheroidal shapes, and inhabit rich galaxy clusters \citep[e.g.][]{2015ApJ...798L..45V}, whereas other UDGs are blue
(g-i$\sim$0.4), have irregular shapes and are found in groups \citep[e.g.][]{2016arXiv161008980R}. Are all these UDGs connected
evolutively? Recently, \citet[][]{2016arXiv161008980R} have suggested a scenario where all this diversity could be understood if UDG
progenitors were born in the field, processed by groups, and ended their lives inhabiting clusters. To answer all the above 
questions and shed more light on the nature of UDGs, it would be extremely useful to have the opportunity to probe, in full detail,
the properties of a close (D$<$15 Mpc) UDG. This would give us the opportunity to explore its individual stars. In particular, it
would be extremely useful to have  some information about the gas content of one of these galaxies. In this work, we present the
serendipitous discovery of a very nearby UDG: the galaxy UGC2162. This galaxy is located in the M77 group (at only 12.3 Mpc distance
from us) and has HI observations. This proximity allows us to have a superb spatial resolution of 60 pc/arcsec. In this work, we
conduct a detailed analysis of the characteristics of this galaxy and confront the observational data with the theoretical
expectations. As we will show, this galaxy is quite rich in HI gas and is currently forming stars at a rate of 0.01 M$_\odot$/yr. If
this galaxy were suddenly depleted of its gas, it would evolve into a red (g-i$\sim$0.8) object with  R$_e$$\sim$1.8 kpc and
$\mu_g(0)$$\sim$27 mag/arcsec$^2$. All of these are characteristics of the population of the faintest UDGs currently found in rich
clusters \citep[][]{2015ApJ...809L..21M,2016ApJ...819L..20B}.

\section{Data}

UGC2162 (R.A.=02h40m23.1s and Dec=+01d13m45s) is located within the IAC Stripe82 Legacy Survey \citep{2016MNRAS.456.1359F}. The
galaxy has a spectroscopic redshift of z=0.00392. The IAC Stripe82 dataset is a careful new co-addition of the SDSS Stripe82 data
with the aim of preserving the faintest surface brightness structures.  The pixel scale of these images is 0.396  arcsec and the
average seeing is 1 arcsec. The following work is based on the rectified images of this dataset
(http://www.iac.es/proyecto/stripe82/). The mean limiting surface brightness of this data co-addition is  29.1, 28.6, and 28.1 mag
arcsec$^{-2}$ in the g, r, and i bands respectively (3$\sigma$ in boxes of 10$\times$10 arcsec). To put this data into context, they
are $\sim$1.2 mag deeper than the Dragonfly images used to explore UDGs in Coma \citep{2015ApJ...798L..45V} and similar to
\citet{2015ApJ...807L...2K}.

UGC2162 is located in the vicinity of M77 (R.A.=02h42m40.7s, Dec=00d00m48s; z=0.00379). Its projected radial
separation to this galaxy is  1.3684 degrees. A redshift independent measurement of the distance to M77
\citep{2009AJ....138..323T} locates this galaxy at a distance of D=12.3($\pm$1.7) Mpc. Because this is the most massive
galaxy of the group, we use its distance as a reliable measurement for the distance of UGC2162\footnote{In what follows, we will consider the uncertainty in the distance to M77 as the main source of error at estimating all the remaining quantities which depend on that distance. These errors will be enclosed within parenthesis to indicate their origin.}. At that distance,
the projected radial separation from UGC2162 to M77 is 293.8($\pm$40.6) kpc and 1 arcsec corresponds to 60($\pm$8) pc. M77 is the central
member of the M77 Group. This is a small group of galaxies that also harbors  NGC 1055, NGC 1073, UGC
2275, UGC 2302, UGCA 44, and Markarian 600.

\begin{figure*}[ht]
\epsscale{1.}
\plotone{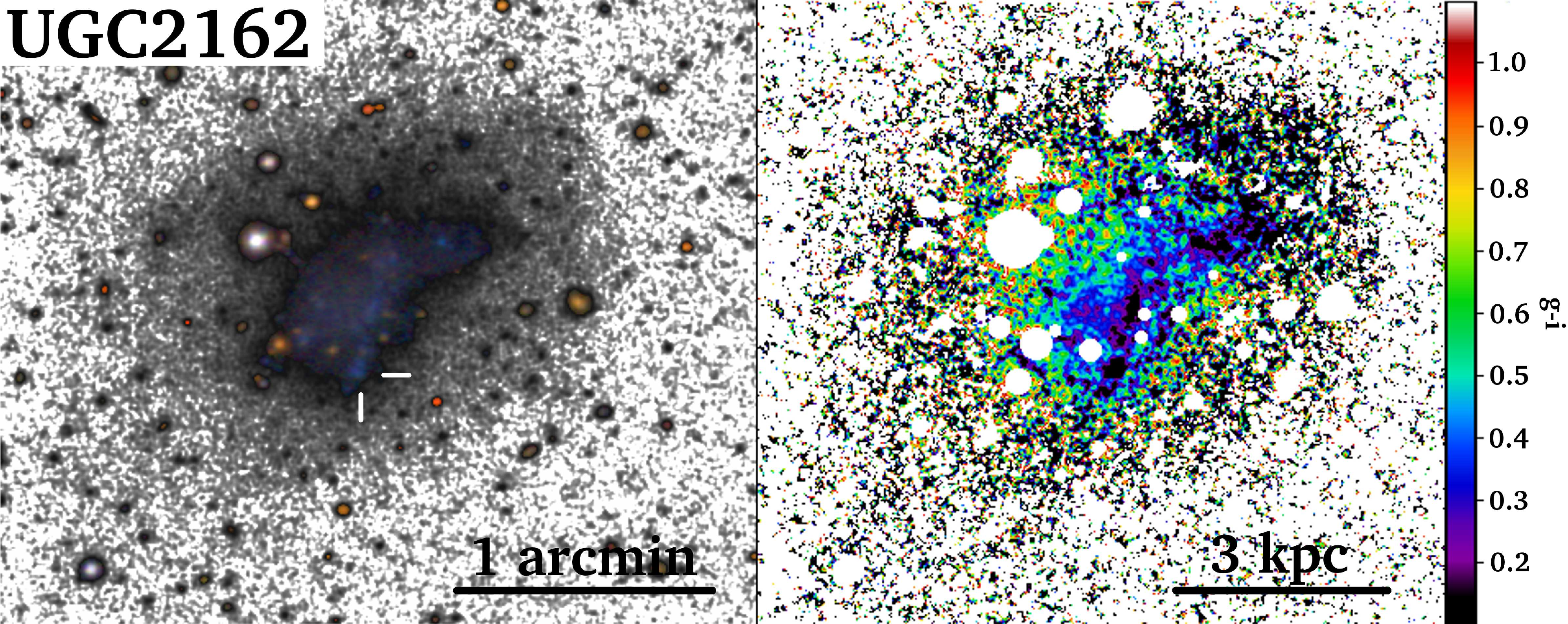}

\caption{{\it Left panel:} g,r,i~ {\it IAC Stripe82} composite image centered on UGC2162. The spatial location of
the SDSS spectrum of this galaxy is indicated with white ticks. {\it Right panel:} g-i
color map of UGC2162. The central irregular region is located on top of a more rounded extended disk-like structure. The white circles are the masked regions used in this work.}

\label{fig:Fig1}
\end{figure*}

Fig. \ref{fig:Fig1} shows a color image of UGC2162 as seen in the IAC Stripe82 images. UGC2162 appears to be an
irregular galaxy, in fact, it has been morphologically classified as Im \citep[][]{1991rc3..book.....D}. The depth
of the Stripe82 image allows us to see that the inner star forming   region of the galaxy is surrounded
by an extended disk-like  structure. 

UGC2162 has been observed by  HIPASS \citep[HI Parkes All Sky Survey;][]{2004MNRAS.350.1195M} with a spectral resolution of 18 km/s.
The HIPASS survey detects at the position of the galaxy a HI line with a radial velocity peak of 1171.9 km/s (in agreement with the
velocity recession of its optical counterpart: 1175$\pm$3 km/s). The HI line flux peaks at 0.089 Jy and has an integrated HI flux
density (S$_{HI}$) of 5.4 Jy km/s. The integrated HI flux density corresponds to a HI mass M$_{HI}$ =
2.36$\times$10$^5$$\times$D$^2$$\times$S$_{HI}$=1.9($\pm$0.6)$\times$10$^8$ M$_\odot$ \citep[see, e.g.][]{2013A&A...558A..18F}. One
can also use the HI line width W$_{20}$=89.7 km/s to infer a dynamical mass M$_{dyn}$ = 2.326$\times$10$^5$ $\times$ (W/2)$^2$
$\times$r$_{HI}$ within the HI radius (r$_{HI}$). To have an estimation of the dynamical mass, it is necessary to correct the line
width W$_{20}$  for the inclination  $i$ of the galaxy, i.e. W= W$_{20}$/$\sin{i}$. We estimate the inclination using the axis ratio
of the 27 mag/arcsec$^2$ isophote (g band). This isophote is still bright enough, but sufficiently far away from the central part of
the galaxy, to produce a reliable estimation of the shape of its outer disk. We obtain an axis ratio b/a=0.7. This translates into an
inclination (under the assumption of a thin disk) of $i$$\sim$45 degrees, and  consequently, W=128 km/s. We assume r$_{HI}$ to be
three times the optical R$_{25}$ radius \citep[see, e.g.][]{2013A&A...558A..18F}. For our galaxy, R$_{25}$=26 arcsec=1.6($\pm$0.3)
kpc (measured in g-band). With these values, we estimate M$_{dyn}$=4.6($\pm$0.8)$\times$10$^9$ M$_\odot$ within the inner R$\sim$5
kpc.

Once a dynamical mass in the inner region of the galaxy has been estimated, it is possible, using the expected cumulative mass curve,
to have an estimation of the total virial mass of the dark matter halo. We follow the same approach as in
\citet[][]{2016ApJ...819L..20B}. In that work, the authors compare the cumulative mass distribution from the EAGLE simulation
\citep[][]{2015MNRAS.451.1247S} with the observed dynamical mass of their galaxy within a given radius (see their Fig. 4, right
panel). From that comparison they infer a virial mass for the dark matter halo. Using our measurement of the enclosed mass
$\sim$4.6($\pm$0.8)$\times$10$^9$ M$_\odot$ within the inner R$\sim$5 kpc, we estimate a virial M$_{200}$ mass similar to that found
by \citet[][]{2016ApJ...819L..20B} for their galaxy (i.e., $\sim$8($\pm$2)$\times$10$^{10}$ M$_\odot$).

\section{Structural and stellar population properties of UGC2162}

To obtain the structural properties of UGC2162 we have used the code IMFIT \citep[][]{2015ApJ...799..226E}. The
surface brightness distribution of the galaxy in each band was modeled using a single S\'ersic component. The S\'ersic model was
convolved with the PSF of the image.  The IAC Stripe82 Legacy Survey provides, for each band, a PSF representative
of the local (0.5$\times$0.5 deg) conditions of the image. To have a first estimate of the spatial coordinates of
the source, the position angle and the effective radius, we use SExtractor. These values are used later as input
parameters for IMFIT. In addition, we mask the closest sources surrounding our galaxy (see Fig. \ref{fig:Fig1}).

We derived the structural parameters of the galaxy in the g, r, and i bands. In all of these bands, the structural parameters of the
galaxy were very similar. We obtained R$_e$=28 arcsec (which is equivalent to 1.7($\pm$0.2) kpc). The central surface brightnesses
were  $\mu_g(0)$=24.4$\pm$0.1 mag/arcsec$^2$, $\mu_r(0)$=24.2$\pm$0.1 mag/arcsec$^2$, and $\mu_i(0)$=24.1$\pm$0.1 mag/arcsec$^2$.
These values have been corrected for Galactic reddening \citep[0.117, 0.081, and 0.060 in the g, r, and i bands
respectively;][]{2011ApJ...737..103S}. The S\'ersic index in all the bands was around n=0.9. The total apparent magnitudes were
g=16.1 mag, r=15.9 mag, and i=15.8 mag.

Using the global color of the galaxy and its absolute magnitude, we can have a rough estimate of its stellar mass.
We follow the recipe by \citet[][]{2015MNRAS.452.3209R} (assuming a Chabrier IMF), using the g-i color and the
absolute magnitude in the r-band (M$_r$=-14.6($\pm$0.3) mag). We obtain a stellar mass of $\sim$2($^{+2}_{-1}$)$\times$10$^7$ M$_\odot$. 

\begin{figure*}[ht]
\epsscale{1.}
\plotone{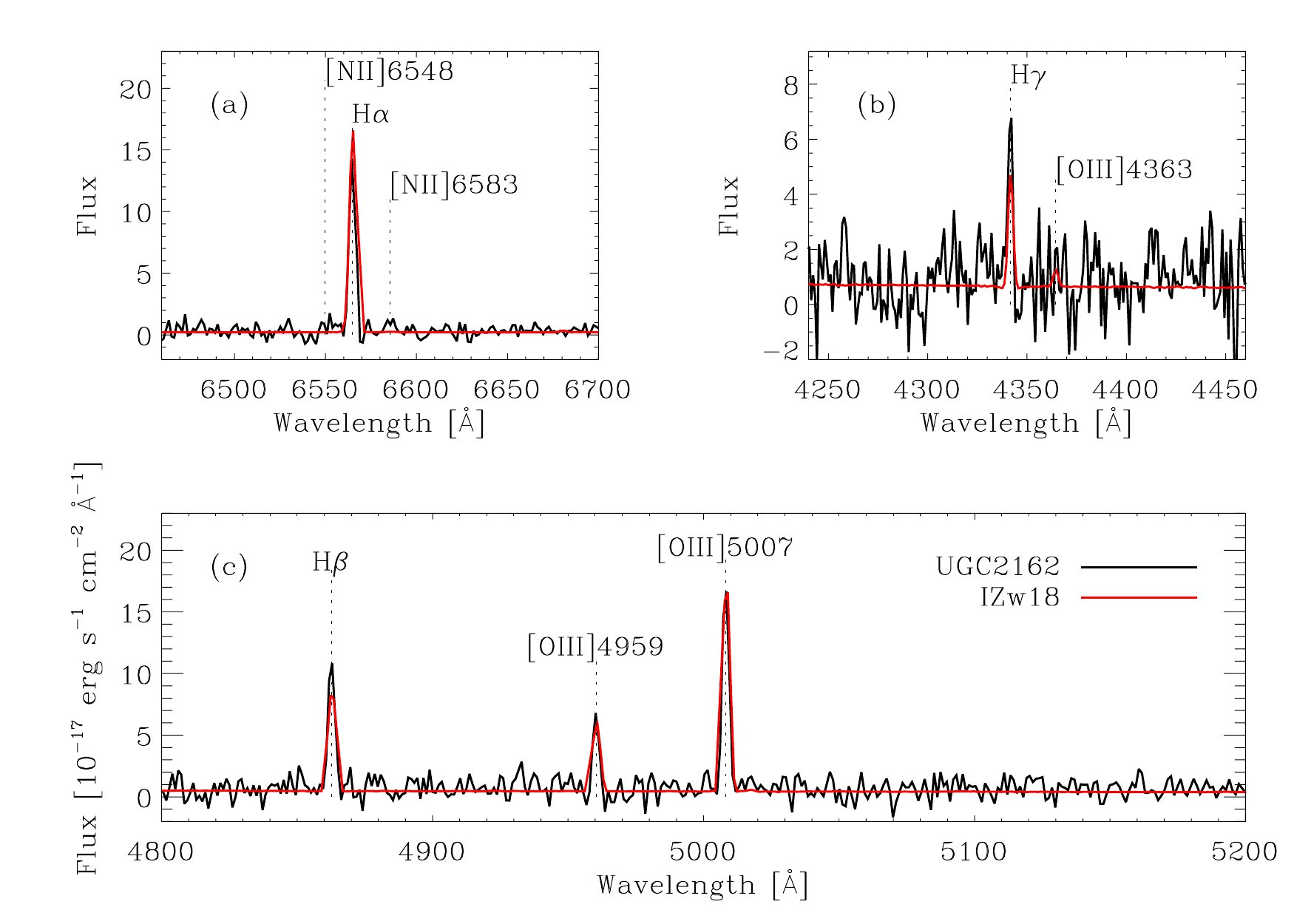}

\caption{Piecewise SDSS-DR12 spectrum of UGC2162 (the black solid line) and of the extremely metal poor galaxy IZw18
(the red solid line), the later included for reference. The main lines are labeled.
(a) Region around H$\alpha$. The ratio between [NII]6583 and H$\alpha$ is used to estimate the
gas-phase metallicity. [NII] is very small in metal-poor systems (see the red solid line).
The flux in H$\alpha$ is used as a proxy for the present SFR. 
(b) Region around [OIII]4363. This line shows up in metal poor systems, but is always very small
(see the red solid line).
(c) Region around [OIII]5007, which includes H$\beta$. The spectrum of IZw18 has been scaled
so that it has the same flux in [OIII]5007 as UGC2162. The units of the flux are shared by the
three panels and are given in panel (c). Wavelengths are in \AA .}

\label{fig:Fig2}
\end{figure*}

UGC2162 has an SDSS spectrum (Plate=1070; Fiber=450; MJD=52591) located at the coordinates: R.A.=40.09751 deg and Dec=1.22476 deg
(see Fig. \ref{fig:Fig2}). The spatial location of the SDSS spectrum is indicated in Fig. \ref{fig:Fig1}. This region corresponds to 
the brightest knot of star formation of the galaxy. This knot has a radius of 1.2 arcsec (Petrosian mag at 90\%) and a magnitude in
the r band of only 21.9 mag. Using the ratio N2$\equiv$[NII]$\lambda$6583/H$\alpha$ from the SDSS spectrum and the calibration by
\citet[][]{2004MNRAS.348L..59P},  we have estimated an oxygen abundance for the star-forming ionized gas 12+log(O/H)=8.22$\pm$0.07,
which corresponds to one-third of the solar abundance. The gas of UGC2162 is fairly metallic for its mass and magnitude, since the
galaxy is a high-metallicity outlier of the mass-metallicity relation and the magnitude-metallicity relation worked out by
\citet[][]{2012ApJ...754...98B}. We also checked the spectrum for the presence of [OIII]$\lambda$4363, which  appears in low
metallicity objects \citep[e.g.][]{2016ApJ...819..110S}. The line is not in the spectrum, which is consistent with the moderate
metallicity inferred from N2.  Since the SDSS spectrum is quite noisy, the estimated O abundance should be regarded as an upper
limit.

Using the H$\alpha$ flux (uncorrected for extinction since H$\beta$ does not seem to be reddened), the distance to
the source, and the recipe of \citet[][]{1998ARA&A..36..189K},  we have estimated the SFR and the surface SFR of
the bright knot, which turn out to be SFR=8.7($\pm$1.2)$\times$10$^{-5}$  M$_\odot$/yr and
$\Sigma_{SFR}$=3.4($\pm$0.5)$\times$10$^{-3}$ M$_\odot$/yr/kpc$^2$. Assuming that the galaxy has 100 such star-forming knots
(reasonable in view of the shape and size of the galaxy) the
total SFR of the galaxy would be SFR=8.7($\pm$1.2)$\times$10$^{-3}$ M$_\odot$/yr. This value is consistent with the value
around 10$^{-2}$ M$_\odot$/yr worked out by \citet[][]{2004AJ....128.2170H} for this object using H$\alpha$
imaging. Using the above value we can derive a specific SFR for UGC2162: sSFR$\sim$5($\pm$0.7)$\times$10$^{-10}$ yr$^{-1}$.
If the SFR of UGC2162 were constant, the galaxy would double its stellar mass in $\sim$2 Gyr.

\section{The future of UGC2162}

UGC2162 is currently located at (a projected separation of) $\sim$300 kpc from M77. Due to its large amount of HI gas ($\sim$10 times
larger than its stellar mass), we can speculate that UGC2162 is undergoing its first infall to the M77 galaxy group. When a galaxy
like UGC2162 falls into a group environment, it suffers a number of physical mechanisms that eventually will quench its star
formation. These mechanisms can be either slow (of the order of a few gigayears) due to gas strangulation \citep[see,
e.g.][]{2008MNRAS.383..593M} or they can be relatively rapid if they are produced by ram pressure stripping
\citep[e.g.][]{2007MNRAS.377.1419W}. Following recent simulations \citep[][]{2015MNRAS.453...14Y}, one can assume that gas rich dwarf
galaxies will be depleted of gas  6 Gyr after their first infall into typical groups of galaxies (10$^{13-13.5}$ M$_\odot$).
Motivated by this number, we simulate how our galaxy would look  in 6 Gyr time if the object were to stop forming stars and followed
a passive evolution. Naturally, this is an oversimplification of the actual evolution of the galaxy, but it can be an interesting
exercise to understand how our object would look  in the future.

\begin{figure*}[ht]
\epsscale{1.}
\plotone{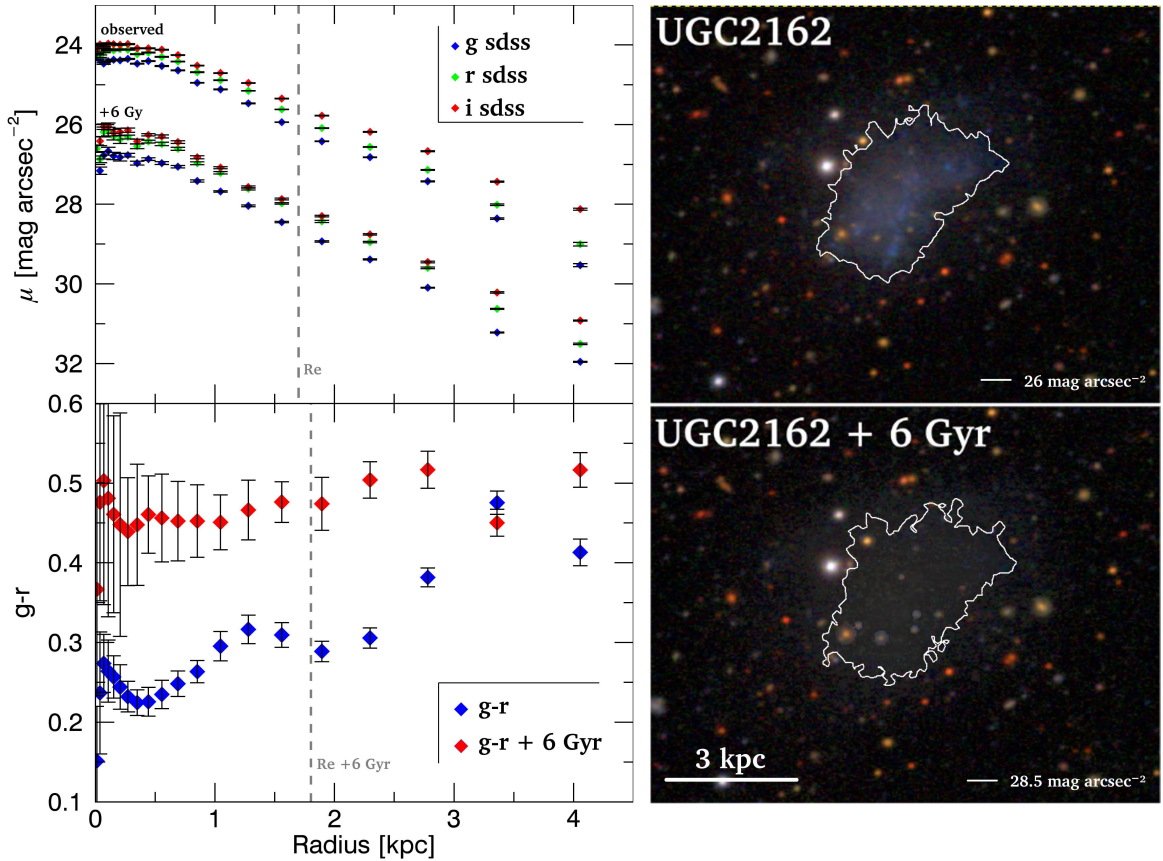}

\caption{{\it Left column:} The g, r, and i present-day surface brightness profiles of  UGC2162  and their passive evolution after 6
Gyr. The lower panel displays the g-r color radial profiles. After 6 Gyr of passive evolution, the galaxy would get significantly
dimmer, redder, and maintain a similar size.  The vertical dashed lines show the position of the effective radius for the present-day
UGC2162 and its potential future evolution. {\it Right column:}  a color composite of how UGC2162 looks today and how the galaxy
would eventually look in the future (after 6 Gyr of passive evolution).  The contours indicate the position of the (g band) 26
mag/arcsec$^2$  (upper panel) and 28.5 mag/arcsec$^2$ isophotes (lower panel).}

\label{fig:Fig3}
\end{figure*}

To model the color and structural evolution of UGC2162, we have used its present-day g-r color map and its current g, r, and i surface
brightness distributions. Then,  we have estimated how every pixel of the images would look  if we make their colors evolve passively
for 6 Gyr. To quantify the color change and the dimming in surface brightness of every pixel, we have used the
\citet[][]{2015MNRAS.449.1177V} models assuming a Kroupa IMF.   Due to this passive evolution, the galaxy would not only change its
global color (becoming g-i=0.77) but  would also get dimmer (by $\sim$2.6 mag/arcsec$^2$). The result of this evolution is
illustrated in Fig. \ref{fig:Fig3}. After 6 Gyr of passive evolution, the galaxy would  have $\mu_g(0)$=27$\pm$0.1 mag/arcsec$^2$ and
R$_e$= 1.8($\pm$0.2) kpc. Its profile shape would not change dramatically, and its passively evolved S\'ersic index $n$$\sim$0.8
would be similar to its original value. With these characteristics, the galaxy would resemble closely the faintest UDG galaxies
discovered so far in the Virgo cluster \citep[][]{2015ApJ...809L..21M}.  In fact, considering the virial mass of UGC2162, in an
eventual future, this galaxy could look very similar to VCC1287, a very low surface brightness UDG galaxy ($\mu_g(0)$=26.7
mag/arcsec$^2$, R$_e$= 2.4 kpc, g-i=0.83, M$_{\star}$$\sim$3$\times$10$^7$ M$_\odot$) inhabiting the Virgo cluster
\citep[][]{2016ApJ...819L..20B}.

\section{Discussion and Conclusions}

The structural and stellar population properties of UGC2162 seem to fit very well within a scheme where the UDGs found both outside
and inside clusters are just different evolutionary stages of the same type of objects \citep[][]{2016arXiv161008980R}. In this view,
UDGs outside clusters would be simply the progenitors of the redder UDGs found in rich clusters. The link between both types of UDGs
would be an evolution due to the removal of their gas produced by the infall of these galaxies in rich environments.  If this picture
is correct UDGs outside dense environments should be dwarf galaxies with a large HI content. In fact, low-surface brightness dwarf
galaxies with a large amount of HI seem to be fairly common in the field \citep[e.g.][]{2013AJ....146....3S,
2015MNRAS.448.2687J,2016ApJ...822..108H,2016arXiv161200273S}. On the contrary, those UDGs found in the richest environments should be
depleted of HI gas due to the removal of this component.

UGC2162 also teaches us an important lesson about UDGs and how these objects are observationally selected. This galaxy has a
relatively low stellar mass ($\sim$10$^7$ M$_\odot$) compared to the general population of UDGs, which peaks at $\sim$10$^8$
M$_\odot$ \citep[e.g.][]{2016arXiv160303494R}. This issue is connected to the way UDGs are defined. UDGs are observationally selected
using their surface brightness ($\mu(0)$$>$24 mag/arcsec$^2$) and size (R$_e$$>$1.5 kpc). This observational definition immediately
biases the selection of the galaxies depending on their stellar populations. The redder UDGs will be more massive and older, whereas
the bluer UDGs will be younger and with lower stellar mass. This implies that if one wants to connect the observed UDGs with their
progenitors or with their descendants, it is important to take this into account. For instance, most of the progenitors of massive
and red UDGs found in rich clusters would not satisfy the observational criteria to be classified as UDGs. These progenitors would
have central surface brightnesses brighter than $\mu(0)$$=$24 mag/arcsec$^2$ and would have been classified as regular (blue) dwarf
galaxies. On the other hand, those blue UDGs that have been discovered outside clusters (as is the case of UGC2162) would evolve into
the less massive (and red) UDGs found in rich clusters (as is the case of VCC1287). Accounting for this selection effect in selecting
UDGs is key to have a comprehensive picture about the nature of these objects and how to connect their different evolutionary stages.

UGC2162 has many of the structural and stellar population properties expected if the large size of this galaxy is the result of
feedback-driven gas outflows \citep[][]{2017MNRAS.466L...1D}. For instance, UGC2162 has a large amount of HI gas and it is currently
forming stars as the cosmological simulations predicted.  In addition, \citet[][]{2017MNRAS.466L...1D} simulations are able to
predict the dwarf-like halo mass of this galaxy, as well as its stellar mass, gas mass, S\'ersic index, effective radius, absolute
magnitude, SFR, irregular appearance, and  off-center-star formation episodes. According to these simulations, these galaxies would
not be at all rare and they will be found in abundance outside clusters. In fact, many of these have  already been detected
\citep[][]{2016arXiv160303494R} in low density environments. If the picture sketched by the cosmological simulations is correct, a
large number of the descendants of UDGs found in low density environments would be found in rich clusters having the following
characteristics: M$_\star$$\gtrsim$10$^7$ M$_\odot$, $\mu_g(0)$$\gtrsim$27 mag/arcsec$^2$, R$_e$$>$1.5 kpc, n$\lesssim$1,
g-i$\sim$0.8 and low HI gas content. These objects would be  hard to find even for current deep surveys. In fact, UDGs with
$\mu_g(0)$$\gtrsim$27 mag/arcsec$^2$ have only be reported by \citet[][]{2015ApJ...809L..21M}, while the remaining UDGs found in rich
clusters have all been found with $\mu_g(0)$$<$27 mag/arcsec$^2$
\citep[e.g.][]{2015ApJ...798L..45V,2015ApJ...807L...2K,2015ApJ...813L..15M,2016A&A...590A..20V,2016arXiv160303494R}. The existence of
a large number of ``dark'' M$_\star$$\sim$10$^7$ M$_\odot$ extended galaxies in rich clusters is a natural prediction of cosmological
simulations if the above evolutionary picture for the UDGs is correct.

Finally, it is worth noting that there are a number of extremely low-surface brightness galaxies at a distance closer than UGC2162
that technically satisfy the criteria to be considered UDGs (i.e. R$_e$$>$1.5 kpc and $\mu(0)$$>$24 mag/arcsec$^2$). These objects
are: a) a satellite of M31, Andromeda XIX \citep[R$_e$$>$1.7 kpc and $\mu(0)$=29.3
mag/arcsec$^2$;][]{2008ApJ...688.1009M,2016ApJ...833..167M}, b) a satellite of our own galaxy, the Sagittarius dwarf \citep[R$_e$=1.6
kpc and $\mu(0)$=25.2 mag/arcsec$^2$;][]{1994Natur.370..194I,2003ApJ...599.1082M} and c) a satellite of the galaxy NGC4449 located at
3.8 Mpc, NGC4449B \citep[R$_e$=2.7 kpc and $\mu(0)$=25.5 mag/arcsec$^2$;][]{2012ApJ...748L..24M,2012Natur.482..192R}. Andromeda XIX
has a very low mass \citep[M$_V$=-9.3 and $\sigma$=4.7 km/s;][]{2013ApJ...768..172C} and it is significantly less massive than the
population of UDGs that has been discussed in the literature (which are a factor of $\sim$100 times more massive). For this reason,
And XIX cannot be considered to be representative of the population of UDGs originally discovered  in the Coma Cluster. In fact, at
that distance, the object would appear invisible in present-day surveys. \citet[][]{2016ApJ...833..167M} discuss the possibility that
the  And XIX large extention could be produced by the gravitational tides of M31. The other two galaxies (though with stellar
masses around 10$^7$ M$_\odot$ or larger) are being tidally disrupted. In this sense, their large effective radii are a consequence
of the ongoing disruption process. Compared to the previous objects, the fact that UGC2162 has a gas reservoir is strong evidence
that it is not diffuse and extended because it is being tidally disrupted. For this reason, UGC2162 is currently the nearest not
tidally disrupted UDG  known, whose large size is probably due to an internal origin alone.

\acknowledgments

We thank the referee for a report. We would also like to thank Michelle Collins for interesting insights into the population of
extremely diffuse Local Group galaxies. We thank Michael Beasley and Chris Brook for their useful comments during the development of
this work. The authors of this paper acknowledge support from grant AYA2013-48226-C3-1-P  from the Spanish Ministry of  Economy and
Competitiveness (MINECO). J.R. thanks the Spanish Ministry of Economy and Competitiveness (MINECO) for financing his PhD through an
FPI grant.





\end{document}